\documentclass{epl}

\title{On the occurrence of oscillatory modulations in the power-law behavior of dynamic and 
kinetic processes in fractals}
\author{M. A. Bab, G. Fabricius and Ezequiel V. Albano.}
\institute{Instituto de Investigaciones Fisicoqu\'{\i}micas Te\'{o}ricas y 
Aplicadas (INIFTA), UNLP, CONICET. Casilla de Correo 16, 
Sucursal 4, (1900) La Plata, Argentina.
}

\begin{document}

\maketitle

\begin{abstract}
The dynamic and kinetic behavior of processes occurring in fractals
with {\it spatial} discrete scale invariance (DSI) is considered. 
Spatial DSI implies the existence of a fundamental scaling ratio ($b_{1}$). 
We address time-dependent physical processes, which as a consequence 
of the time evolution develop a characteristic 
length of the form $\xi \propto t^{1/z}$, where $z$ 
is the dynamic exponent. So, we conjecture that the interplay between the physical process
and the symmetry properties  of the fractal leads to the occurrence of {\it time} DSI
evidenced by soft log-periodic
modulations of physical observables, with a fundamental time scaling ratio
given by $\tau = b_{1}^{z}$. The conjecture is tested numerically 
for random walks, and representative systems of broad 
universality classes in the fields of irreversible and equilibrium critical phenomena.     
\end{abstract}

The understanding of the kinetic and dynamic behavior of the wide variety of phenomena 
and processes, taking place in disordered and low-dimensional media, is a challenging issue 
of increasing interdisciplinary interest \cite{ency,katya}. The simplest and paradigmatic 
example is, most likely, the effort devoted to the study of random walks \cite{havlin}. 
However, many other examples can also be quoted such as the dynamics of critical 
systems in condensed matter physics \cite{katya,haye}; 
chemical reactions in catalysts, porous media, nano- and microcavities; annihilation 
reactions of a wide diversity ranging from magnetic monopoles in the early stages of the 
Universe to excitons in polymeric matrices \cite{kop}; epidemic propagation 
of diseases and forest
fires in far from-equilibrium systems \cite{ency,haye,ips}; 
coarsening dynamics in many systems involving 
fluids, magnets, and eventually the formation of opinion in social systems \cite{haye}; etc.
In this letter we address the subtle interplay between time-dependent processes and the 
symmetry  properties of the underlying structure or substrate where the considered 
process actually takes place. For this purpose, fractal media that 
exhibit discrete scale invariance (DSI) \cite{ref3} are considered. DSI is a weak kind of 
scale invariance such that an 
observable $O(x)$, 
obeys the scaling law 
\begin{equation}
O(x)= \mu(b) O(bx),  
\label{ecdsi}
\end{equation}
\noindent under the change $x \rightarrow bx$. Here $b$ is no longer an arbitrary 
real number, as in the case of \textit{continuous} scale invariance, but it can only 
take specific discrete values of the form $b_n =(b_1)^n$, 
where $b_1$ is a fundamental scaling ratio. If an observable $O(x)$ satisfies 
equation (\ref{ecdsi}) 
for an arbitrary $b$, it necessary has to obey a power law of the type 
$O(x) = Cx^\alpha$, where $\alpha $ is an exponent. But in the case of DSI, 
the solution of equation (\ref{ecdsi}) yields
\begin{equation}
O(x) = x^\alpha F\left( \frac{\log (x)}{\log (b_1)}\right),  
\label{ecdsi2}
\end{equation}
\noindent where $F$ is a periodic function of period one.
The detection of soft oscillations in spatial domain \cite{ref3}
is a signature of spatial DSI.
Let us consider physical processes that develop a time-dependent  characteristic
length $\xi(t)$ that increases monotonically,
and analyze the behavior of kinetic or dynamic observables, $O(t)$, 
which characterize these physical processes. 
If a biunivocal $t = f(\xi )$ relationship holds, we conjecture $\tilde{O}(\xi )=O(t(\xi))$
to obey spatial DSI of the form
\begin{equation}
\tilde{O}(\xi ) = \mu (b_1 ^n) \tilde{O}(b_1 ^n \xi ) .
\label{sdsi}
\end{equation}
Now, if we assume that $\xi \propto t^{1/z}$, where $z$ is a dynamic exponent,
it is straightforward to show that $O(t)$ has to obey time DSI. 
In fact, by using equation (\ref{ecdsi2}) for $\tilde{O}(\xi )$ and replacing $\xi$
by its explicit time dependence, one has
\begin{equation}
O(t) =C t^{\alpha /z} F\left( \phi + \frac{\log (t)}{\log (b_1 ^z)}\right),  
\label{ecdsi3}
\end{equation}
where $C$ and $\phi$ are constants.
So, if our conjecture given by equation (\ref{sdsi}) 
is correct, we would expect to obtain  
a logarithmic periodic modulation of time observables characterized by a
time-scaling ratio $\tau$ given by
\begin{equation}
\tau = b_1 ^z.
\label{app2}
\end{equation}

A relationship between spatial and time DSI was implicitly suggested in 
connection with the behaviour of 
earthquakes on a pre-existing hierarchical fault structure \cite{eplsornette} and by ourselves
in order to account for the dynamic behaviour of a magnet on a fractal substrate \cite{we2}. 
This relationship has also been established in the mathematical literature for the case 
of a random walk on Sierpinski graphs \cite{mate}.
In this letter we propose the conjecture given by equation(\ref{sdsi}) to obtain an
explicit relationship between spatial and time DSI according to equation(\ref{app2}). 
Furthermore, we test our conjecture for three paradigmatic cases in detail, namely 
i) the behavior of a single random walk and the diffusion-controlled reaction among random 
walkers \cite{havlin,kop}, 
ii) the contact process as an archetype of an epidemic process 
exhibiting irreversible critical behavior \cite{haye,ips,jan}, and 
iii) the Ising model that represents a broad universality of reversible 
(equilibrium) critical phenomena \cite{ising}.

\begin{figure}
 \centerline{\includegraphics[width=6.0cm,angle=-90]{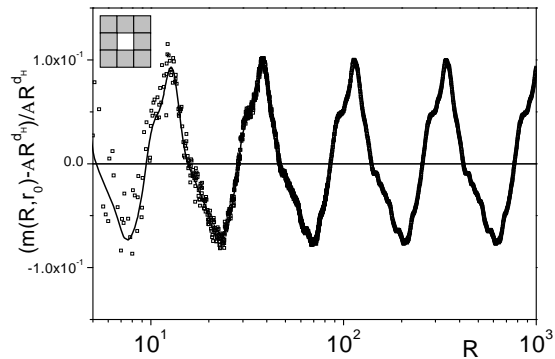}}
 \caption{Plot of the number of particles of the fractal located  
          within a distance $R$ from the origin $r_{o}$ $(m(R,r_{o}))$, 
          as a function of $R$.
          Notice that the vertical axis has been normalized by using the 
          results of the standard power-law fit with an exponent given
          by the fractal dimension $d_{F} = ln(8)/ln(3)$. 
          For the oscillation, the fit (shown by a continuous line) yields a frequency given by
          $\omega = 2\pi/log(3)$, and involves up to the $5^{th}$ harmonic.
          The generating cell of the SCIR(9,8) is shown in the top-left corner.}
 \label{Fig1}
\end{figure}

While equations (\ref{ecdsi3}) and (\ref{app2}) are expected to hold 
for processes occurring in all substrate exhibiting spatial DSI,
for the sake of simplicity in this paper we have simulated physical 
processes on Sierpinski Carpets (SC), which have 
both infinite and finite ramification orders (IR and FR, respectively),
and fractal dimensions within the interval $1.631 \le d_{F} \le 1.975$. 
In order to build up a generic SC (SCIR(b,c) or SCFR(b,c)), 
a square in $d=2-$dimensions is segmented into b$^d$ subsquares 
and $c$ of them are then removed, and this segmentation 
process is iterated on the remaining subsquares a number $k$ of 
segmentation steps. 
DSI may become evident by measuring some topological 
property of the fractal, such as the average number of 
sites belonging to the fractal as a function of the distance to the origin \cite{sorDLA},
e.g. see figure \ref{Fig1}.

\begin{figure}
 \centerline{\includegraphics[width=7.0cm]{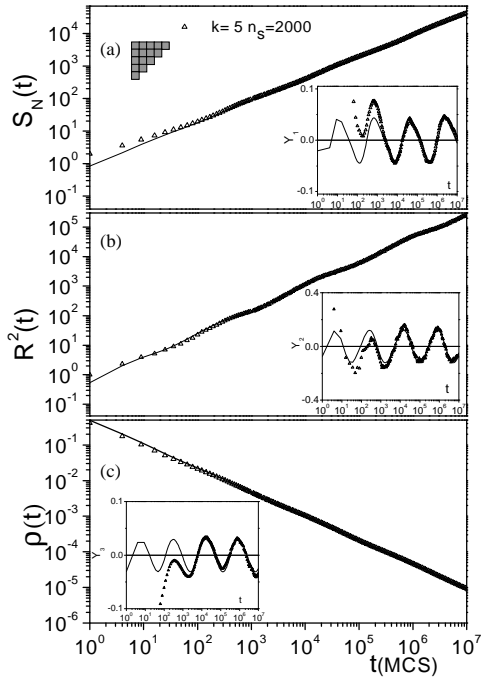}}
\vskip -0.20 true cm 
 \caption{Log-log plot of (a) $S_{N}(t)$, (b) $R^{2}(t)$, and (c) $\rho(t)$ versus $t$,
          respectively. Results obtained for random walks on the SCFR(5,10) shown
          in (a). Each of the insets shows, on a linear-log scale, the oscillatory 
          modulation of the corresponding main panel, after previous subtraction 
          and subsequent normalization by the fitted power law. These quantities are
          denoted by $Y_{m}, m = 1, 2$ and $3$, as taken form the upper to the lower panel,
          respectively.}
\label{Fig2}
\end{figure}

\begin{figure}
\includegraphics[height=8cm,width=5cm,angle=-90]{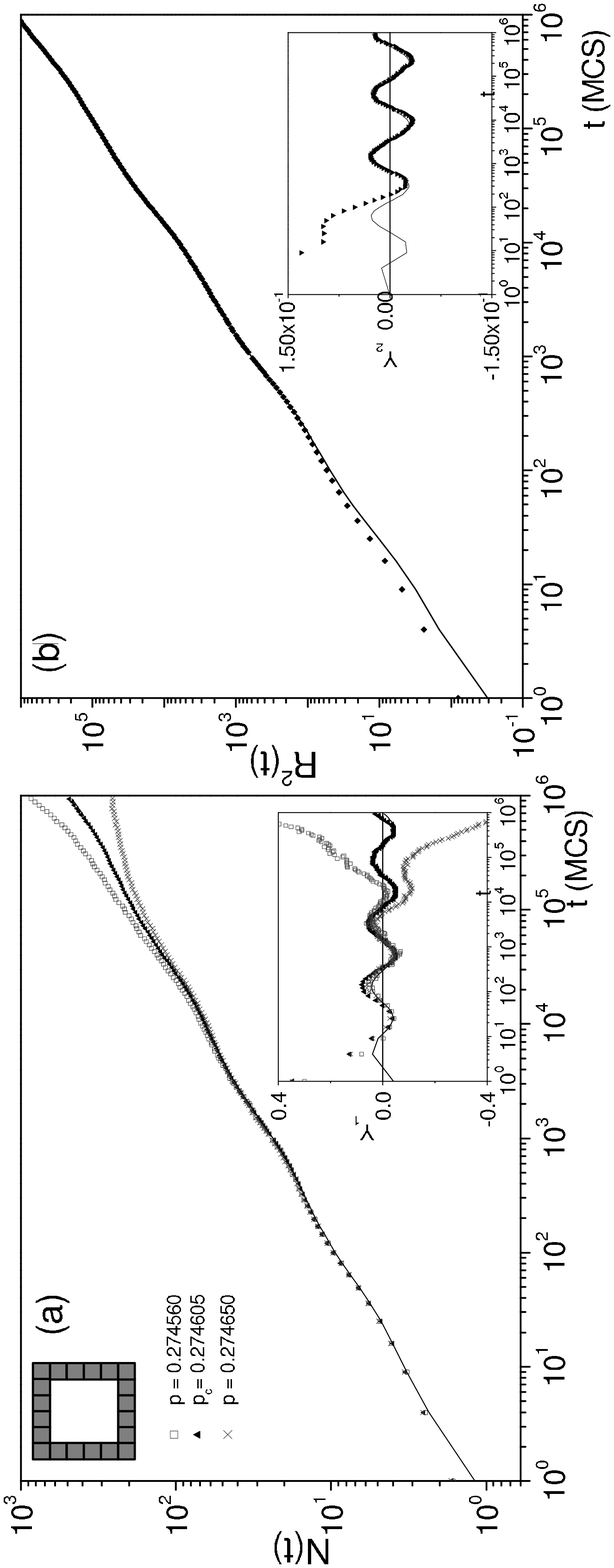} 
\vskip -0.10 true cm
\caption{Data corresponding to the CP. 
(a) Log-log plots of the number of active sites versus time obtained for different
values of $p$. In the left-top corner we show the generating 
cell. (b)  Log-log plots of $R^{2}$ versus time obtained at criticality.  
The insets show the modulating function.}
\label{fig4}
\end{figure}

Relevant kinetic observables of {\it random walks} are the average 
number of distinct sites visited after $N$ steps ($ S_{N} $) and the 
mean square displacement from the origin  ($ R^{2} $), given by 
\begin{equation}
S_{N}(t) \propto  t^{d_{s}/2}, \hspace{0.5cm}  {\rm and} \hspace{0.5cm}  R^{2} (t)\propto  t^{\nu}, 
\label{SN}
\end{equation}
\noindent respectively \cite{havlin,kop}. Here we have taken $N \propto t$, while 
$d_{s}$ is the spectral dimension and $\nu$ is the random walk exponent.
For homogeneous media in dimensions $d \ge  2$ one 
has $d_{s} = 2$ and $\nu = 1$ leading to 
classical (standard) diffusion. However, for $d < 2$ one has $d_{s} < 2$, and 
the diffusion is anomalous, as diffusion controlled annihilation 
reactions (such as e.g. $ A + A \rightarrow  0 $) that no longer obey 
the classical textbook equation for a second-order reaction, but instead are 
described by the following anomalous rate equation \cite{kop}
\begin{equation}
\frac{d \rho(t)}{dt} = - k_{o} \rho(t)^{X}, \rho \rightarrow 0 ,
\label{AR}
\end{equation}
\noindent  where $\rho $ is the density of reacting walkers, $k_{o}$ is the
rate constant, and the anomalous reaction 
order is given by $ X = 1 + 2/d_{s}$ \cite{kop}.
Figures \ref{Fig2}(a), (b), and (c) show log-log plots of $S_{N}$, $R^{2}$ 
and $\rho$ versus $t$ (notice that after a simple 
integration, equation (\ref{AR}) yields $\rho \propto  t^{-d_{s}/2}$), respectively.
Results are averaged over $n_{s}$ different configurations.
The soft oscillatory modulations of the power-law behavior of these 
observables become more evident after 
a proper normalization,
as shown in the insets of figure \ref{Fig2}.
Table I summarizes the results obtained by fitting the data 
with equation (\ref{ecdsi3}) up to the $2^{nd}$ harmonic. 
An excellent agreement (within error bars)
between the relevant parameters is obtained: the exponent $d_{s}$, 
evaluated for a single walk, agrees with the result corresponding to 
the annihilation reaction; the period of the 
oscillation is independent of the observable as is the exponent $z$ 
obtained by applying equation (\ref{app2}). 
Furthermore, by comparing equations (\ref{ecdsi3}) and (\ref{SN}),
and taking $R^{2}(t) \propto \xi(t)^{2}$, it follows
that $z = 2/\nu$, so that the evaluation of $R^{2}(t)$ provides an independent
test for the conjectured relationship given by equation (\ref{app2}),
(see the $5^{th}$ column of Table I).
\begin{table}
\caption{Results obtained for random walks by 
fitting the data shown in figure \ref{Fig2} by means of 
equation (\ref{ecdsi3}). The dynamic exponents are 
obtained by using equation (\ref{app2})
with $b_{1} = 5$. More details in the text.}
\begin{tabular}{cccccccc}
\hline\hline
Observable & Exponent           & log($\tau) $  &   $z$    & $2/\nu$     \\ \hline\hline
$S_{N}(t)$ & $d_{s}$ = 1.342(4) & 1.785(15) & 2.55(2)  &  ---        \\ 
$\rho(t)$  & $d_{s}$ = 1.350(8) & 1.72(3) & 2.46(4) &  ---        \\ 
$R^{2}(t)$ & $\nu$ = 0.809(3)   & 1.75(4) & 2.50(6)      & 2.47(1)     \\ 
\hline\hline
\end{tabular}
\end{table}%
The {\it contact process} (CP) \cite{haye,ips,jan}, is a model for the spreading of an 
epidemic. It is a one-component non-equilibrium lattice model with spontaneous 
annihilation and autocatalytic creation of particles. In the CP each site is either 
vacant or 
occupied by a single particle. Particles are annihilated at a rate $p$, independent of 
the states of other sites, and vacancies become occupied at a rate $ M/Z $, where $M$ is the 
number of occupied nearest-neighboring (NN) sites, and $Z$ is the total number 
of NN sites. The vacuum state of the CP is absorbing 
and it is reached for $p$ exceeding a critical threshold ( $p \ge p_{c}$).
For $p < p_{c}$ the state is fluctuating with an stationary propagation of the 
activity. So, just at $p_{c}$ one has a second-order irreversible phase transition    
(IPT). IPT´s are often 
characterized by means of dynamic (epidemic) measurements that allow avoiding 
undesired finite-size effects \cite{haye,jan}. 
Epidemic measurements are started with a configuration very 
close to the absorbing state, e.g. the vacuum state with a single particle in the 
center of the sample for the case of the CP \cite{jan}. 
Subsequently, one measures the average 
number of particles ($N(t)$) at time $t$, the probability that the system has not 
entered into the absorbing state ($P(t)$) at time $t$, and the average mean-square 
distance of spreading ($R^{2}(t)$). In 
order to obtain appropriate statistic one has to perform many epidemic runs (typically we  
performed $ N_{e} \approx 4\times 10^{4}$ different runs).
From the scaling Ansatz of IPT´s it follows that, at criticality, the quantities 
defined above 
obey, in nonfractal $d-$dimensional spaces, power-law behaviors of the form
\begin{equation}
N(t) \propto t^{\eta}, (a)~~
P(t) \propto t^{-\delta}, (b)~~
{\rm and}~~
R^{2}(t) \propto t^{z_{e}}, (c)
\label{epi}
\end{equation}
\noindent where $\eta $, $\delta$, and $ z_{e} $ are the critical exponents.
(Notice that, for the sake of clarity, we have used $\nu$ in equation (\ref{SN}) 
and $z_{e}$ (equation (\ref{epi}(c)) for the 
exponent of the mean square displacement of the randon walk and the 
epidemics, respectively.)
Figures \ref{fig4} (a) and (b) show that log-log plots of $N(t)$ and $R^{2}(t)$ 
versus $t$ exhibit soft log-periodic oscillations 
modulating the power-law behavior given by equations (\ref{epi})(a) and (c),
respectively. By fitting $N(t)$ 
with a modulated power law of the type given by equation (\ref{ecdsi3}), we can 
easily identify the critical threshold and decouple the oscillations, 
as shown in the inset of figure \ref{fig4}(a). Also, for $ R^{2}(t)$, 
with the aid of equation (\ref {epi})(c) we obtained 
$ z_{e} $ as listed in Table II. 
As in the case of the random walk, one has that  
$z = 2/z_{e} $, so that the epidemic study allows us to perform independent measurements 
of the dynamic exponent, see Table III. By considering the error bars,
all results corresponding to the CP, which are summarized in Table II, show
excellent agreement: the period is independent of the observable, as are 
the values of $z$, and the independent measurements of $z$ are consistent.

\begin{table}
\caption{Results obtained by performing epidemics in the fractals
listed in the first column. The logarithmic period, and the dynamic
exponents obtained by using equation (\ref{app2}), for the different
observables, are listed in subsequent columns.}
\begin{tabular}{cccccccc}
\hline\hline
$SCIR$ & $log(\tau)_{N}$  & $z$ & $ log(\tau)_{P}$ & $z$ & $log(\tau)_{R^2}$ & 
$z$ & $2/z_{e} $   \\ \hline\hline
(3,1)  & 0.882(9)  & 1.85(2)  & ---     & ---      & 0.87(2)    & 1.83(4)  & 1.91(1) \\ 
(4,4)  & 1.122(4)  & 1.863(7) & 1.18(2) & 1.96(3)  & 1.12(2)    & 1.86(4) & 1.867(5) \\ 
(6,16) & 1.415(16) & 1.82(2)  & 1.41(2) & 1.81(2)  & 1.42(2)   & 1.83(2)  & 1.835(5) \\ 
\hline\hline
\end{tabular}
\end{table}%

{\it The Ising model} is a lattice system where each site $i$ is occupied by a 
two-state spin variable $ \sigma_{i} = \pm 1 $\cite{ising}. In the absence of external 
magnetic fields, the Hamiltonian ($ H $) is given by
$H = - J \sum_{i,i' = 1 }^{ N } \sigma_{i} \sigma_{i'}$,
where $ J > 0 $ is the coupling constant and the summation runs over 
nearest neighbor sites. In  $ d > 1 $ dimensions the Ising model exhibit a 
continuous phase transition between a ferromagnetic and a paramagnetic phase, 
which has become the archetypical example for the study of critical phenomena. 
By starting from a fully ordered configuration with magnetization $ M = 1$, 
corresponding to the ground state at $ T = 0 $, after quenching to the critical point one 
observes a relaxation of the form $M(t) \propto t^{-\beta/ \nu z}$,
where $\beta$ and $\nu $ are the order parameter and correlation length critical 
exponents, respectively. We have shown that the relaxation of $M(t)$ exhibits
clear evidence of a log-periodic modulation, in the case of the SCIR(4,4) \cite{we} 
and SCIR(3,1) \cite{we2}. 
We now assume that the observed behavior can be understood in terms of time DSI,
allowing us to evaluate $z$. Furthermore, we have performed additional simulations 
for different fractals, which are summarized in Table III. 
\begin{table}
\caption{Critical exponent $\frac \beta {\nu z}$, oscillation logarithmic period
and dynamic exponent $z$ for the Ising model on several Sierpinski Carpets.}
\label{tab??} 
\begin{tabular}{cccccc}
\hline\hline
SC & $\frac \beta {\nu z}$ & $log (\tau)$ & $z$
\\ \hline\hline
SCIR(5,1)  & 0.0508(4)  & 1.61(4) & 2.30(6)\\
SCIR(3,1)  & 0.0341(1)  & 1.28(3) & 2.69(6)\\
SCIR(4,4)  & 0.0110(1)  & 2.16(1) & 3.59(1)\\
SCIR(5,9)  & 0.00214(5) & 3.47(2) & 4.97(3)\\ \hline\hline
\end{tabular}
\end{table}

Summing up, we have well established a relationship between discrete scaling 
symmetry invariance, as it is present in  
fractal structures, and a corresponding
symmetry in the time domain for the kinetic and dynamic evolution of 
different physical systems defined 
in such structures. We conjecture that the relationship can be obtained by making 
time observables satisfy spatial DSI symmetry, when they are expresed
in terms of the growing length, which characterizes the time evolution. 
We have also shown that our conjecture implies the presence of time 
logarithmic oscillations of period $\tau$, linked to the fundamental scaling ratio of the 
fractal ($b_1$) and the dynamical exponent ($z$) (equation (\ref{app2})).
For the sake of simplicity, we have tested numerically our conjecture 
for relevant archetypical cases on SC substrates. However, our
arguments are valid for any fractal exhibiting spatial DSI.


Acknowledgments: This work is supported financially by 
CONICET, UNLP, and ANPCyT (Argentina). GF acknowledge the ICTP
for working facilities.

\end{document}